\documentclass[aps,pra,reprint,showpacs,superscriptaddress,nofootinbib]{revtex4-2}
\usepackage{amsmath,amssymb,amsfonts}
\usepackage{graphicx}
\usepackage{dcolumn}
\usepackage{epstopdf}
\usepackage[T1]{fontenc}
\usepackage{newtxtext}
\usepackage{newtxmath}
\usepackage{subfigure}
\usepackage{float}
\usepackage{bm}
\usepackage{natbib}

\usepackage[unicode]{hyperref}
\hypersetup{
	unicode=true,          
	a4paper=true,
	plainpages=false,
	pdftitle={Title of PDF},    
	pdfauthor={Author of PDF},     
	pdfsubject={Subject of PDF},   
	colorlinks=true,       
	linkcolor=blue,          
	citecolor=blue,        
	filecolor=blue,      
	urlcolor=blue           
}
\newcommand{\refeq}[1]{Eq.~(\ref{#1})}

\newcommand{\reffig}[1]{Fig.~\ref{#1}}

\DeclareMathAlphabet{\mathcal}{OMS}{cmsy}{m,b}{n,it}

\begin{document}

\title{Polarization-controlled pattern formation in antiparallel dipolar binary condensates}

\date{\today}

\author{Zhijun Zhang}
\thanks{These authors contributed equally to this work.}
\affiliation{College of Physics, Nanjing University of Aeronautics and Astronautics, Nanjing 211106, China}
\affiliation{Key Laboratory of Aerospace Information Materials and Physics (Nanjing University of Aeronautics and Astronautics), MIIT, Nanjing 211106, China}

\author{Weijing Bao}
\thanks{These authors contributed equally to this work.}
\affiliation{College of Physics, Nanjing University of Aeronautics and Astronautics, Nanjing 211106, China}
\affiliation{Key Laboratory of Aerospace Information Materials and Physics (Nanjing University of Aeronautics and Astronautics), MIIT, Nanjing 211106, China}

\author{Changjian Yu}
\affiliation{College of Physics, Nanjing University of Aeronautics and Astronautics, Nanjing 211106, China}
\affiliation{Key Laboratory of Aerospace Information Materials and Physics (Nanjing University of Aeronautics and Astronautics), MIIT, Nanjing 211106, China}

\author{Jinbin Li}
\affiliation{College of Physics, Nanjing University of Aeronautics and Astronautics, Nanjing 211106, China}
\affiliation{Key Laboratory of Aerospace Information Materials and Physics (Nanjing University of Aeronautics and Astronautics), MIIT, Nanjing 211106, China}

\author{Gentaro Watanabe}
\email{gentaro@zju.edu.cn}
\affiliation{School of Physics and Zhejiang Institute of Modern Physics, Zhejiang University, Hangzhou, Zhejiang 310027, China}
\affiliation{Zhejiang Province Key Laboratory of Quantum Technology and Device, Zhejiang University, Hangzhou, Zhejiang 310027, China}

\author{Kui-Tian Xi}
\email{xiphys@nuaa.edu.cn}
\affiliation{College of Physics, Nanjing University of Aeronautics and Astronautics, Nanjing 211106, China}
\affiliation{Key Laboratory of Aerospace Information Materials and Physics (Nanjing University of Aeronautics and Astronautics), MIIT, Nanjing 211106, China}

\begin{abstract}
	We investigate non-equilibrium pattern formation in an antiparallel two-component dipolar Bose–Einstein condensate by varying the polarization angle and the trap aspect ratio. At finite tilt, the condensate supports stripe order. Quenching the angle to zero triggers a roton-assisted, mushroom-like corrugation that destroys translational order and drives the system into labyrinthine textures, whereas a slow linear ramp produces long-lived curved stripes that ultimately converge to labyrinths. Population imbalance strongly biases the evolution: the minority component preferentially fragments into a stable droplet array while the majority remains comparatively diffuse; once formed, the droplet crystal is robust under polarization hysteresis with largely reversible shape changes and unchanged lattice topology. The trap aspect ratio controls both the initial stripe number and the instability timescale, with tighter axial confinement accelerating corrugation and yielding denser labyrinths at late times. All behaviors arise within a quasi-two-dimensional mean-field regime where beyond-mean-field corrections are negligible; accordingly, the droplets reported here are not self-bound in free space. The observed textures (such as stripes, curved stripes, and labyrinths) mirror the taxonomy and instability pathways of nuclear “pasta” morphologies (rods and slabs) known from neutron-star and supernova matter, highlighting polarization angle, trap geometry, and population imbalance as practical, experimentally accessible controls for selecting and steering patterns in dipolar mixtures.
\end{abstract}

\maketitle

\section{Introduction}\label{sec:intro}

Dipolar quantum gases composed of atoms with magnetic dipole moments, such as chromium \cite{griesmaier2005}, dysprosium \cite{minglu2011}, and erbium \cite{aikawa2012}, have drawn considerable attention in both theoretical and experimental research. Prominent phenomena include quantum droplets stabilized by quantum fluctuations \cite{Ferrier2016,edmonds2020,schmitt2016self,Baillie2018}, exotic phases of matter like honeycombs, triangles, and stripes \cite{Mukherjee2023,Roccuzzo2019,zhang2019,Zhang2021,Aitor2022,Bland2022,lihui2024,Mukherjee2023super,Gallem2022,sanchez2023ring,Young2022,Tanzi2019,Hertkorn2021,Ripley2023}, and quantized vortices emerging during rotational dynamics \cite{Gallem2020,klaus2022,Ancilotto2021}.

A dipolar Bose-Einstein condensate (dBEC) exhibits both isotropic contact interactions and anisotropic, long-range dipole-dipole interactions. This combination leads to novel quantum phases and complex dynamical properties.
Studies of two-component dBECs have revealed diverse phenomena: 
pattern formation (e.g., frog, mushroom, and labyrinthine structures) at the condensate interface during instability dynamics \cite{xikuit2018}; 
supersolid phases emerging without quantum fluctuations \cite{LiShaoxiong2022};
quantum droplet phases \cite{Smith2021,BissetR2021,Sinha2023,Scheiermann2023catalyza};
immiscible binary supersolids \cite{Blandt2022};
complex droplet patterns modulated by scattering lengths \cite{Halder2023}. 
Additionally, vortex lattice structures form under rotating magnetic fields \cite{Kumar2017,kumar2018vortex,Silva2023}, and binary mixtures of atomic species (e.g., erbium-dysprosium) have been experimentally realized \cite{trautmann2018,Politi2022,Durastante2020}.

The attractive or repulsive nature of dipolar interactions in dBECs is governed by tilting the dipole polarization using an external magnetic field \cite{Giovanazzi2002,Baillie2020}.
This polarization critically determines phase separation and enables vortex lattice formation \cite{Kumar2019,Aleksandrova2024,Staudinger2023}. 
Furthermore, experimental observations in dBECs include hexagonal lattices, self-organized stripe states, and distinct ground states \cite{HalderS2022,Wenzel2017,Lepers2018,Ghosh2022,Mishra2020}.

Recent studies of two-component dBECs with strong dipole-dipole interactions (DDI) reveal diverse phenomena including ring lattice states, dual superfluids, and binary supersolids induced by competition between intra-/inter-component contact interactions and DDI \cite{zhangyongc2024,HalderDas2024Induced}. 
Systems featuring opposite magnetic dipole moments have garnered particular attention, and the key findings include: observation of droplet toroidal and crystalline structures in antiparallel dipole condensates, especially when both components share identical scattering lengths; discovery of a curved stripe phase (distinct from conventional straight stripes) induced by quantum fluctuation repulsion \cite{ArazoGallem2023}.

In this paper, we investigate a two-component dipolar Bose-Einstein condensate of chromium atoms with antiparallel magnetic dipole moments. By systematically tuning the polarization angle and the trapping potential, we observe the emergence of stripe phases and quantum multi-droplet states. The stripe phase progressively evolves into curved, wavy patterns (resembling wavy domain structures in classical magnetic fluids) and eventually forms intricate labyrinthine configurations. We further explore the impact of population imbalance, demonstrating the robust stability of quantum multi-droplet states under hysteretic dynamics. Moreover, we show that the trap aspect ratio critically influences the number of stripes and promotes the formation of mushroom-like and labyrinthine structures. 
The stripe, curved-stripe, and labyrinthine textures reported here echo the ``nuclear pasta'' morphologies [e.g. slabs (``lasagna''), rods (``spaghetti''), cylindrical bubbles, and related patterns] predicted for subnuclear densities in neutron-star matter and observed in quantum-molecular-dynamics simulations of compression and phase transitions in supernova cores \cite{WatanabePRL2005,WatanabePRL2009,CaplanRMP2017}. 
While the microscopic interactions and thermodynamic constraints differ, the common thread is pattern selection by competing interactions and geometric frustration.

This paper is organized as follows. In Sec. \ref{sec:formulation}, we formulate the problem. In Sec. \ref{sec:dynamics}, we investigate the nonlinear dynamics of stripe phases by quench (Sec. \ref{subsec:quench}) and linearly decrease (Sec. \ref{subsec:linear_decrease}) the polarization angle, as well as hysteresis behavior (Sec. \ref{subsec:hysteresis}). In Sec. \ref{sec:trap_ratio}, we show stationary pattern formation as the trap ratio is varied. In Sec. \ref{sec:nuclear_pasta}, we discuss the analogy to nuclear ``pasta'' and frustrated pattern formation. In Sec. \ref{sec:conclusion}, we conclude the study.

\section{Formulation}\label{sec:formulation}

A two-component Bose-Einstein condensate (BECs) with dipolar interactions can be described using the non-local Gross-Pitaevskii (GP) equations:

\begin{align}\label{gp}
	i \hbar \frac{\partial}{\partial t} \Psi_{i} \left(\bm{r}\right)
	= \Bigg[ & -\frac{\hbar^2}{2m} \nabla^2 + V \left( \bm{r} \right) + \sum_{j = 1}^{2} g_{ij} \big| \Psi_{j}\left(\bm{r}\right) \big|^2 \nonumber\\
	&+ \sum_{j = 1}^{2} \int U_{ij} \left( \bm{r} - \bm{r'} \right) \big| \Psi_{j} \left( \bm{r'} \right) \big|^2 d \bm{r'} \Bigg] \Psi_{i}\left(\bm{r}\right),
\end{align}
where $m$ is the atom mass, $\Psi_{1}\left(\bm{r}, t\right)$ and $\Psi_{2}\left( \bm{r}, t\right)$ represent the three-dimensional wave functions of the two components, and the normalization condition is $\int |\Psi_{i}|^2 d \bm{r} = N_{i}$, with $i = 1, 2$ and $N_{i}$ is the number of atoms in each component. 
The right-hand side of \refeq{gp} represents, in order, the kinetic energy term, the external potential field, the contact interactions, and the dipole-dipole interactions(DDIs) of the system. The external harmonic potential is
\begin{equation}
	V\left(\bm{r} \right) = \frac{1}{2} m \left[ \omega_{\perp}^{2} ( x^{2} + y^{2} ) + \omega_{z}^{2} z^{2} \right],
\end{equation}
where $\omega_{\perp}$ and $\omega_{z}$ denote the radial frequency in the $x$-$y$ plane and the axial frequency along the $z$-axis, respectively. 
The contact interaction coupling constant in \refeq{gp} is
\begin{equation}
	g_{ij} = \frac{4 \pi \hbar^2 a_{ij}}{m},
\end{equation}
where $a_{12}$ ($a_{21}$), $a_{11}$ ($a_{22}$) denote the inter-particle and intra-particle scattering lengths, respectively. The DDI has the form
\begin{equation}\label{ddi}
	U_{ij} \left( \bm{r} \right) = \frac{\mu_{0} \mu_{i} \mu_{j} \left[ 1 - 3 \left( \hat{d} \cdot \hat{\bm{r}} \right)^{2} \right]}{4 \pi r^{3}},
\end{equation}
where $r = |\bm{r}|$ is the distance between two dipoles, $\mu_{0}$ is the vacuum permeability, and $\mu_{i}$ and $\mu_{j}$ are the magnetic dipoles of an atom in each component. The vector $\hat{d}$ on the right-hand side of \refeq{ddi} represents the dipole polarization direction
\begin{equation}
	\hat{d} = \cos \alpha \hat{z} + \sin \alpha \hat{x},
\end{equation}
where $\alpha$ denotes the angle between the dipole polarization direction and the $z$-axis direction. If the polarization direction is along the $ z $-axis then we have $\alpha = 0$, $\hat{d} = \hat{z}$, $\left( \hat{d} \cdot \hat{\bm{r}} \right)^{2} = \left( \hat{z} \cdot \hat{\bm{r}} \right)^{2}$, and \refeq{ddi} can be rewritten as $U_{ij} \left( \bm{r} \right) = \gamma \mu_{0} \mu_{i} \mu_{j} ( 1 - 3 \cos \theta^{2} ) / 4 \pi r^{3}$, where $\theta$ is the angle between the directions of polarization and the distance $r$ between the two dipoles. We selected two components, both composed of $^{52}$Cr atoms, with magnetic dipole moments of $\mu_1 = 6\mu_B$ and $\mu_2 = -6\mu_B$, respectively, where $\mu_B$ is the Bohr magneton. This two-component dipolar Bose-Einstein condensate (BEC) system can be implemented experimentally using the $^{7}S_{3}$ states with $m_{J} = -3$ and $+3$ states of $^{52}$Cr \cite{griesmaier2005,Santos2006}. The frequencies of the external harmonic trapping potential are $(\omega_{\perp}, \omega_{z}) = 2\pi \times (100, 800)$ Hz. 

We consider the two components with densities $n_1 (\bm{r})$ and $n_2 (\bm{r})$, and define the in-phase (density) $n_d(\bm{r})$ and out-of-phase (spin) $n_s(\bm{r})$ as 
\begin{align}
	n_d (\bm{r}) &= \frac{1}{\sqrt{2}} \left[ n_1 (\bm{r}) + n_2 (\bm{r}) \right], \\
	n_s (\bm{r}) &= \frac{1}{\sqrt{2}} \left[ n_1 (\bm{r}) - n_2 (\bm{r}) \right].
\end{align}

The contact interaction energy is given by 
\begin{align}
	E_c &= \frac{1}{2} \int d^3 r \left[ g_{11} n_1^2 (\bm{r}) + 2g_{12} n_1 (\bm{r}) n_2 (\bm{r}) + g_{22} n_2^2 (\bm{r})  \right] \nonumber\\
	&= \frac{1}{2} \int d^3 r \left[ (g + g_{12}) n_d^2 (\bm{r}) + (g - g_{12}) n_s^2 (\bm{r}) \right],
\end{align}
with $g_{11} = g_{22} \equiv g$.

The dipolar interaction energy is given by 
\begin{align}
	E_{dd} &= \frac{1}{2} \int d^3 r d^3 r^{\prime} U(\bm{r} - \bm{r}^{\prime}) \left[ n_1 (\bm{r}) - n_2 (\bm{r}) \right] \left[ n_1 (\bm{r}^{\prime}) - n_2 (\bm{r}^{\prime}) \right] \nonumber\\
	&= \int d^3 r d^3 r^{\prime} U(\bm{r} - \bm{r}^{\prime}) n_s (\bm{r}) n_s (\bm{r}^{\prime}),
\end{align}
with $U(\bm{r} - \bm{r}^{\prime})$ is the dipolar interaction.
The dipolar interaction couples only to the spin (out-of-phase) branch and cancels identically in the density (in-phase) branch.

Performing Fourier transforms gives the momentum-dependent kernels in the total energy:
\begin{align}
	V_d(\bm{k}) &= g + g_{12}, \\
	V_s(\bm{k}) &= g - g_{12} + 2U(\bm{k}).
\end{align}
Thus, the density branch, which controls the compressibility and provides the dominant zero-point contribution, is purely contact and $\bm{k}$-independent. Hence, any dipolar Lee-Huang-Yang (LHY) contribution via $V_d(\bm{k})$ is absent, while all dipolar anisotropy resides in the spin branch through $U(\bm{k})$. For the symmetric contact case ($g = g_{12}$), the spin stiffness $V_s(\bm{k})$ is small over the stable background, so its zero-point contribution is parametrically suppressed in the three-dimensional case. Therefore, the net LHY correction is negligible in our regime, and we omit it without loss of accuracy.

The three-dimensional nonlocal Gross-Pitaevskii equations (\ref{gp}) are solved via pseudo-spectral methods with fast Fourier transforms (FFTs). The scaling feature of \refeq{gp} is characterized by the following dimensionless parameters: $\omega_{x} / \omega_{\perp}$, $\omega_{z} / \omega_{\perp}$, $4 \pi N_{i,j} a_{i,j} / l$,  $N_{i,j} \mu_{0} \mu_{i} \mu_j/ (4 \pi \hbar \omega_{\perp} l^3)$ with $l = \sqrt{ \hbar / m \omega_{\perp} }$.

\section{DYNAMICS OF STRIPE PHASES}\label{sec:dynamics}

\begin{figure}[t]
	\begin{center}
		\includegraphics[width=0.5\textwidth]{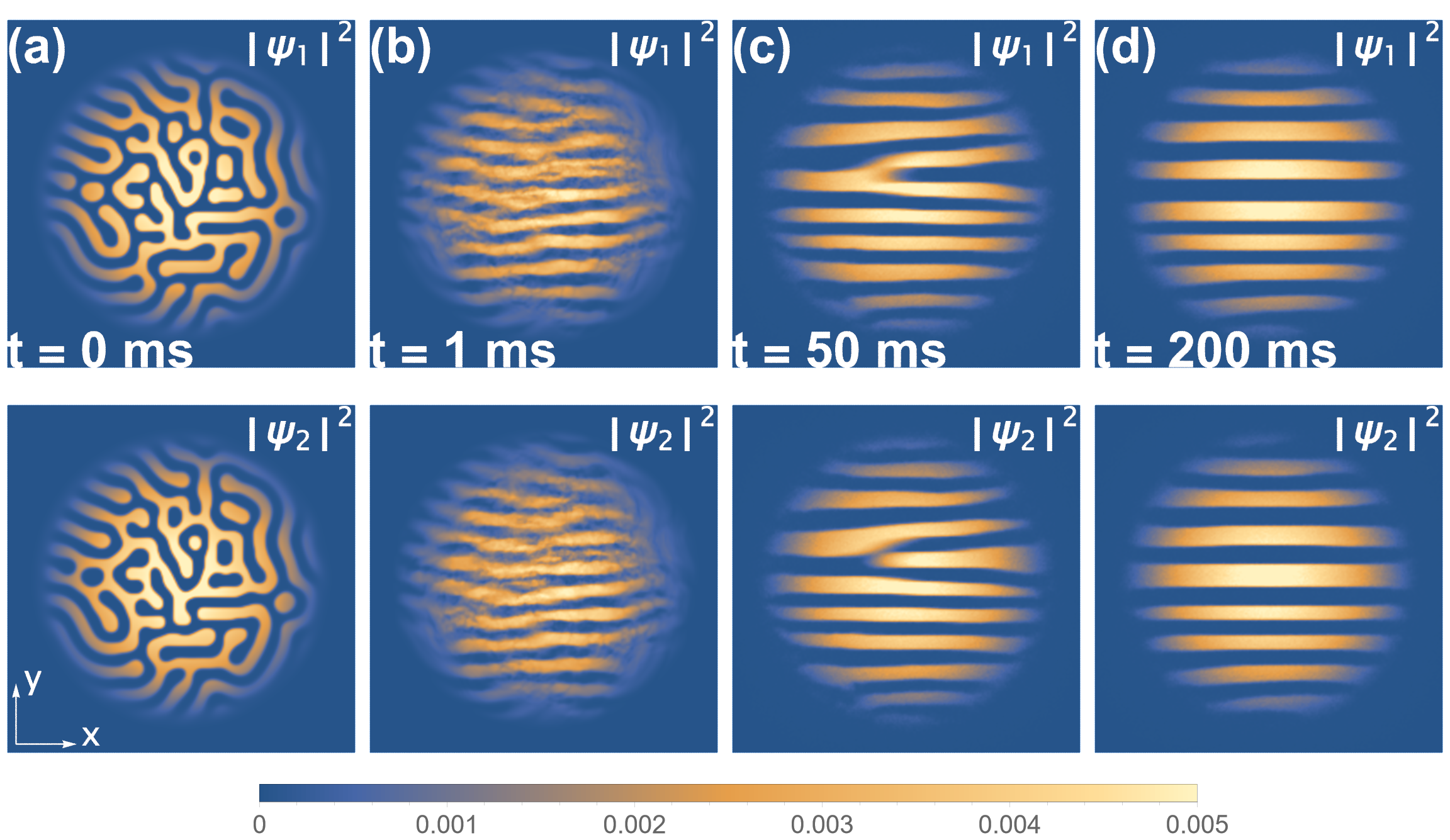}
	\end{center}\vspace{-0.5cm}
	\caption{\label{fig:n1_effect_angle_pi3}
		Column-density profiles of the two components during real-time evolution. 
		(a) With zero polarization angle ($\alpha = 0$), both components display a labyrinthine pattern. 
		(b)--(d) For a finite angle $\alpha = \pi/3$, the textures evolve toward stripes. 
		Parameters: $a_{11} = a_{22} = a_{12} = 100 a_B$, $N_1 = N_2 = 2 \times 10^{6}$, $(\omega_{\perp}, \omega_{z}) = 2 \pi \times (100, 800)$ Hz, and $\mu_1 = -\mu_2 = 6\mu_B$.}
\end{figure}

We consider a two-component dipolar BEC confined in a pancake-shaped harmonic potential with $ \left( \omega_{\perp}, \omega_{z} \right) = 2\pi \times \left( 100, 800 \right) $ Hz. 
A stationary state at zero polarization angle ($\alpha = 0$) is obtained via imaginary-time propagation with dipole-dipole interactions included. 
This state is then used as the initial condition for real-time dynamics, where at $t = 0$ ms the polarization angle is suddenly set to $\alpha = \pi / 3$. In experiment, the polarization axis can be steered optically by exploiting the vector (fictitious) magnetic field generated by elliptically polarized, far-detuned light; its magnitude and orientation follow the beam polarization, enabling all-optical control of $\alpha$. Moreover, by choosing wavelengths/polarizations that address the two components differently (e.g., different atomic species with distinct vector polarizabilities), one can independently engineer the effective field seen by each component---realizing different magnitudes, orientations, or temporal protocols with minimal cross-talk \cite{GRIMM200095}.
The resulting column-density distributions  (integrated along $z$) of each component are shown in \reffig{fig:n1_effect_angle_pi3}. 
Throughout, we fix the intra- and interspecies scattering lengths to $ a_{11} = a_{22} = a_{12} = 100 a_B $ and the atom numbers to $ N_{1} = N_{2} = 2 \times 10^6 $. 

For $\alpha = 0$, both components display a labyrinthine pattern [\reffig{fig:n1_effect_angle_pi3} (a)]. Upon introducing the finite polarization angle, the condensate evolves from a labyrinth to a stripe texture [Figs.~\ref{fig:n1_effect_angle_pi3} (a)--(d)], and the stripe order becomes increasingly robust during the subsequent dynamics.

\subsection{Quench of the polarization angle}\label{subsec:quench}

\begin{figure*}[t]
	\begin{center}
		\includegraphics[width=\textwidth]{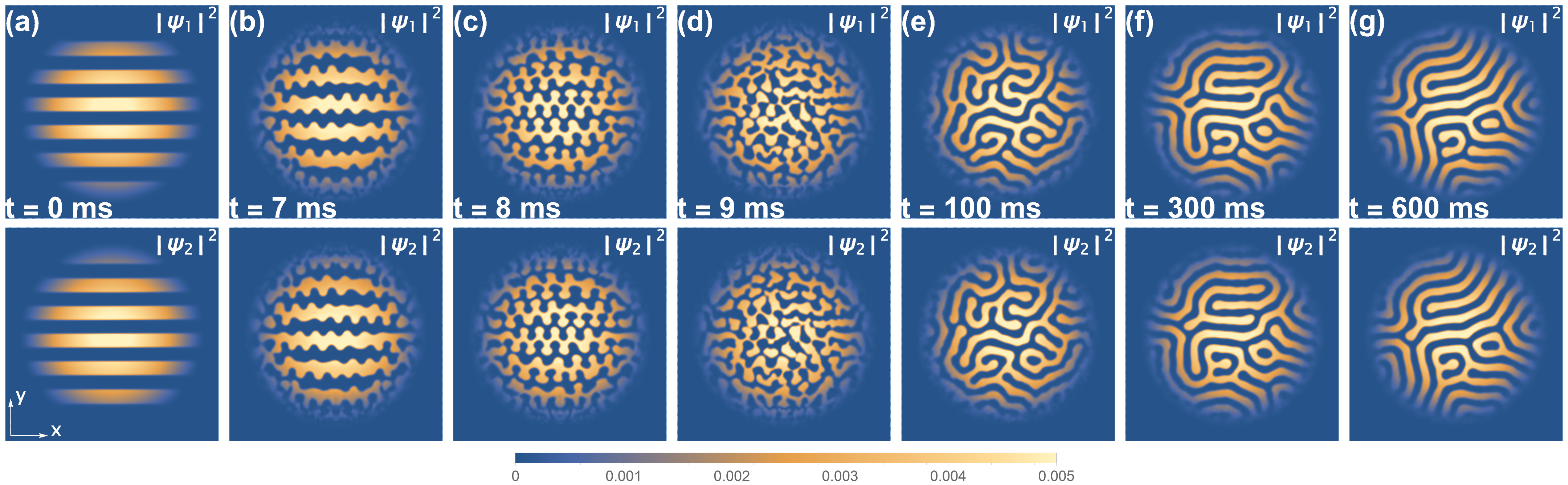}
	\end{center}\vspace{-0.5cm}
	\caption{\label{fig:quench_pi3}
		Column-density profiles $|\psi_{1}|^{2}$ and $|\psi_{2}|^{2}$ during real-time dynamics for $N_1/N_2=1$.
		(a) At $\alpha=\pi/3$, both components realize a stripe phase.
		At $t=0$ ms the polarization angle is quenched to $\alpha=0$ and then held fixed; the two-component stripe order destabilizes, developing mushroom-like protrusions and complex labyrinthine textures [panels (b)–(g)].
		Parameters: $N_{1}=N_{2}=2\times10^{6}$, $a_{11}=a_{22}=a_{12}=100 a_B$, $(\omega_{\perp}, \omega_z)=2\pi\times(100, 800)$ Hz, $\mu_{1}=6\mu_{B}$, and $\mu_{2}=-6\mu_{B}$.}
\end{figure*}

\begin{figure*}[t]
	\begin{center}
		\includegraphics[width=\textwidth]{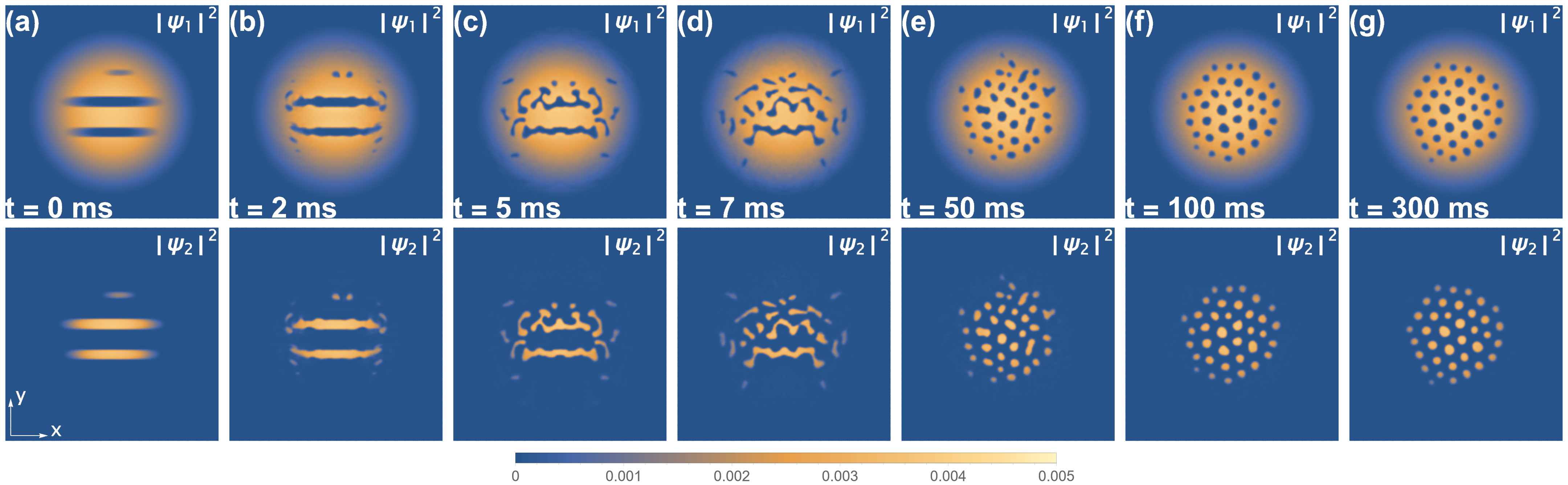}
	\end{center}\vspace{-0.5cm}
	\caption{\label{fig:n6_pi3_quench}
		Column-density profiles $|\psi_{1}|^{2}$ and $|\psi_{2}|^{2}$ during real-time dynamics for a population-imbalanced mixture with $N_1/N_2=6$ ($N_1=2 \times 10^{6}$).
		(a) Initial state at $\alpha=\pi/3$.
		At $t=0$ ms, the angle is quenched to $\alpha=0$ and held thereafter.
		Following the quench, the stripe order rapidly fragments into discrete, droplet-like density peaks that subsequently stabilize [panels (b)–(g)].
		Other parameters as in \reffig{fig:quench_pi3}.}
\end{figure*}

We probe the non-equilibrium response by tuning the polarization angle away from a stripe ground state and monitoring the ensuing real-time dynamics. 
Unless stated otherwise, column densities are integrated along $z$, and parameters are as in \reffig{fig:quench_pi3}.

We first prepare equal populations ($N_1=N_2=2 \times 10^6$) at $\alpha=\pi/3$, where both components display wide, regular stripes with a well-defined orientation [\reffig{fig:quench_pi3} (a)]. 
At $t=0$ ms we suddenly quench the angle to $\alpha=0$ and keep it fixed thereafter; representative snapshots are shown in \reffig{fig:quench_pi3}. 
The subsequent evolution exhibits three qualitatively distinct stages. 
(i) \emph{Early time.} On a few-millisecond timescale, transverse corrugations nucleate on top of the straight stripes; by $t \simeq 8$ ms clear mushroom-like protrusions are visible on essentially every stripe [\reffig{fig:quench_pi3} (c)]. 
(ii) \emph{Intermediate time.} Corrugations grow and merge, severing the stripes into meandering filaments and short segments; orientational order is rapidly lost and domains of competing orientations proliferate [see Figs.~\ref{fig:quench_pi3} (d) and (e)].
(iii) \emph{Late time.} The condensate relaxes toward a labyrinthine morphology with a broad distribution of local orientations [Figs.~\ref{fig:quench_pi3} (f)--(g)], consistent with the $\alpha=0$ textures reported in Ref.~\cite{xikuit2018}. 

To assess the role of intercomponent coupling and relative densities, we repeat the same quench for an imbalanced mixture with $N_1/N_2=6$ ($N_1=2 \times 10^{6}$); see \reffig{fig:n6_pi3_quench}. 
Whereas both components stripe at $\alpha=\pi/3$ [\reffig{fig:n6_pi3_quench} (a)], the quench to $\alpha=0$ renders the state unstable and the minority component rapidly fragments into discrete droplets [Figs.~\ref{fig:n6_pi3_quench} (b)--(g)], which then stabilize with little subsequent coarsening. 
The majority component remains comparatively diffuse and does not form a commensurate droplet lattice on the simulated timescales. 
Operationally, we find that the droplet number saturates around $t = 100$ ms while the mean nearest-neighbor spacing changes by less than a few percent thereafter, indicating the system has become dynamically arrested in a metastable droplet configuration.

\subsection{Linear decrease of the polarization angle}\label{subsec:linear_decrease}

\begin{figure*}[t]
	\begin{center}
		\includegraphics[width=\textwidth]{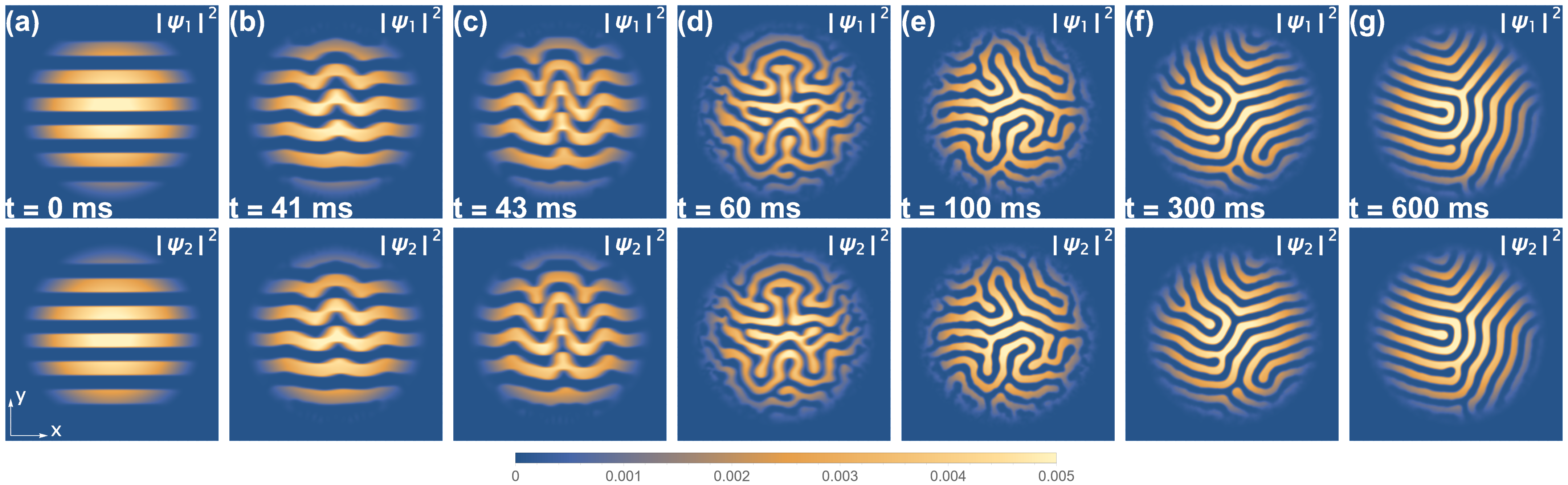}
	\end{center}\vspace{-0.5cm}
	\caption{\label{fig:linear_pi3}
		Column-density profiles $|\psi_{1}|^{2}$ and $|\psi_{2}|^{2}$ for a linear ramp of the polarization angle.
		(a) The system is initialized at $\alpha=\pi/3$.
		The angle is then reduced linearly to $\alpha=0$ over $100$ ms and held at zero thereafter.
		During the ramp, the stripe-axis symmetry breaks and the stripes bend from their centers, generating progressively intricate patterns [panels (a)–(e)].
		Once $\alpha=0$ [panels (e)–(g)], the textures evolve toward a labyrinthine state.
		Parameters: $N_{1}=N_{2}=2 \times 10^{6}$, $a_{11}=a_{22}=a_{12}=100 a_B$, $(\omega_{\perp}, \omega_z)=2 \pi \times(100,800)$ Hz, $\mu_{1}=6\mu_{B}$, and $\mu_{2}=-6\mu_{B}$.}
\end{figure*}

\begin{figure*}[t]
	\begin{center}
		\includegraphics[width=\textwidth]{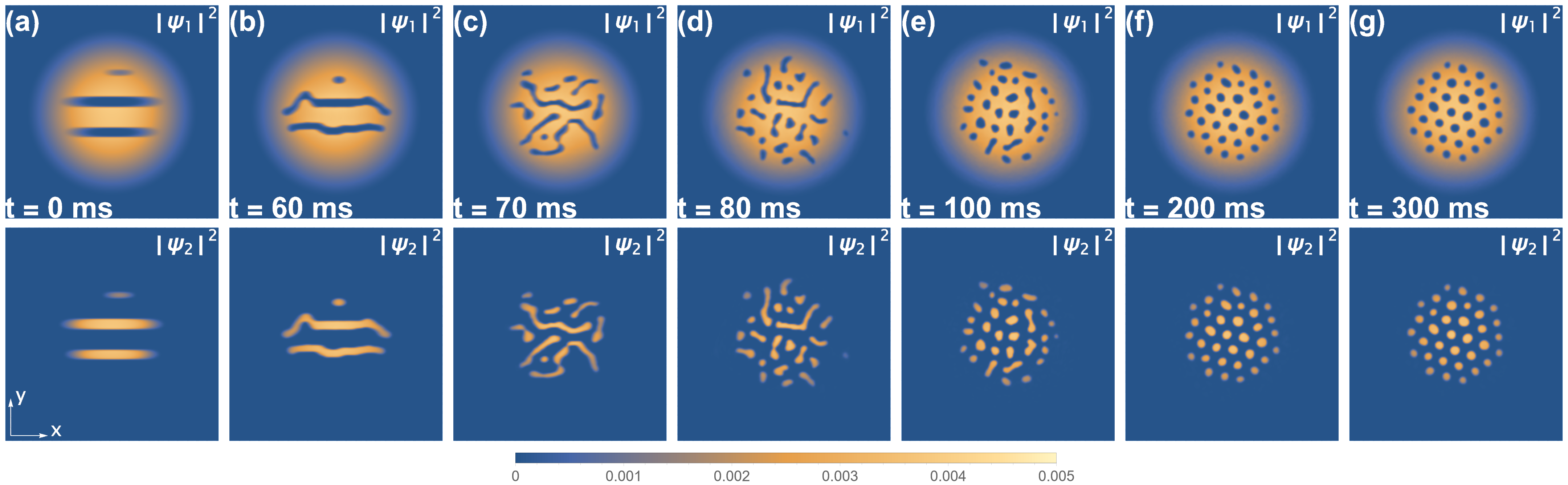}
	\end{center}\vspace{-0.5cm}
	\caption{\label{fig:n6_pi3_linear_100ms}
		Column-density profiles $|\psi_{1}|^{2}$ and $|\psi_{2}|^{2}$ for a linear ramp with population imbalance $N_1/N_2=6$ ($N_1=2 \times 10^{6}$).
		(a) Initial configuration at $\alpha=\pi/3$; (a)–(e) $\alpha$ is ramped linearly to $0$ over $100$ ms; (e)–(g) $\alpha=0$ is then held constant.
		Other parameters are as in \reffig{fig:linear_pi3}.}
\end{figure*}

We next steer the dynamics with a slow protocol that linearly decreases the polarization angle from $\alpha=\pi/3$ to $0$ over $100\,\mathrm{ms}$ and then holds it fixed. 
Explicitly,
\[
\alpha(t)=
\begin{cases}
	\pi/3\,(1-t/\tau_{\rm ramp}), & 0\le t\le \tau_{\rm ramp}=100\,\mathrm{ms},\\[2pt]
	0, & t>\tau_{\rm ramp}.
\end{cases}
\]
Representative snapshots are shown in \reffig{fig:linear_pi3} (balanced) and \reffig{fig:n6_pi3_linear_100ms} (imbalanced) for $a_{11}=a_{22}=a_{12}=100 a_B$ and $(\omega_{\perp}, \omega_z)=2 \pi \times (100,800)$ Hz.

Starting from stripes at $\alpha=\pi/3$, the global stripe axis first loses alignment and individual bands bend from their centers, generating an axisymmetric, curved-stripe texture [Figs.~\ref{fig:linear_pi3} (b) and (c)]. 
As the ramp proceeds, curvature increases, stripe segments reconnect, and multi-directional domains proliferate; once $\alpha=0$, the pattern relaxes toward a labyrinthine morphology [Figs.~\reffig{fig:linear_pi3} (e)--(g)].

With an imbalance of $N_1 / N_2 = 6$, the same ramp delays but does not prevent fragmentation: the minority component’s stripes deform noticeably by $t \simeq  60$ ms and then separate into individual droplets that remain long-lived after the ramp ends at $\alpha = 0$ [Figs.~\ref{fig:n6_pi3_linear_100ms} (b)--(g)]. 
The majority component stays comparatively diffuse on the simulated timescales and exhibits a void structure formed by being displaced by the minority component.
Lowering $\alpha$ progressively reduces the in-plane anisotropy of the dipolar kernel, weakening the directional bias that stabilizes straight stripes. 
The ramp shifts the roton-like wave vector and reduces transverse stiffness, enhancing the susceptibility to long-wavelength splay or undulation modes. 
This yields a robust sequence—curved stripes, meanders and to labyrinths—consistent with the curved-stripe intermediates reported in Ref.~\cite{ArazoGallem2023}.

Relative to the quench (Sec.~\ref{subsec:quench}), the linear ramp injects energy more gradually and keeps the system in a curved-stripe metastable manifold for longer [cf.\ \reffig{fig:quench_pi3} (b) vs \reffig{fig:linear_pi3} (b)]. 
Nevertheless, both protocols converge to labyrinthine textures once $\alpha=0$, because removal of anisotropy eliminates the preferred modulation direction and unlocks competing domains.

\subsection{Hysteresis dynamics}\label{subsec:hysteresis}

\begin{figure}[t]
	\begin{center}
		\includegraphics[width=0.5\textwidth]{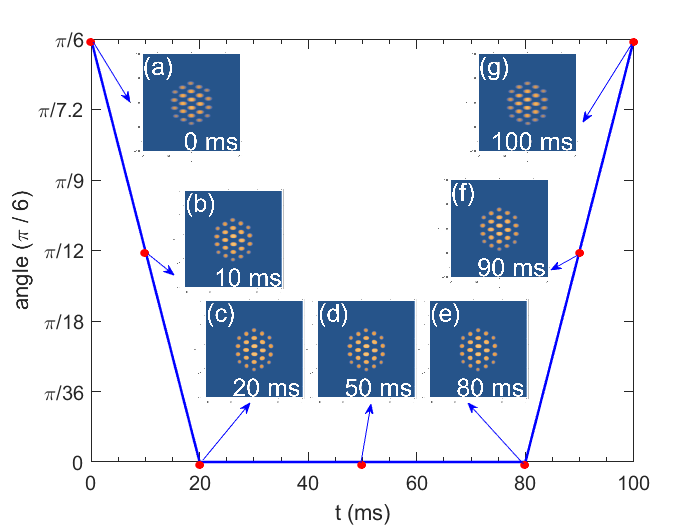}
	\end{center}\vspace{-0.5cm}
	\caption{\label{fig:n6_pi6_dynamic20ms}
		Column-density distributions of component 2, $|\psi_{2}|^{2}$, for $N_1/N_2=6$ under a hysteresis protocol.
		(a) Ground state at $\alpha=\pi/6$ and $t=0$ ms shows a hexagonal array of trap-bound droplets.
		(a)–(c) The angle is ramped linearly to $0$ within $20$ ms; (c)–(e) $\alpha=0$ is held until $80$ ms; (e)–(g) the angle is ramped back to $\pi/6$ by $t=100$ ms.
		Other parameters: $a_{11}=a_{22}=a_{12}=100 a_B$, $(\omega_{\perp}, \omega_{z})=2 \pi \times (100,800)$ Hz, $\mu_1=6\mu_B$, and $\mu_2=-6\mu_B$.}
\end{figure}

We now examine how repeated linear adjustments of the polarization angle affect an already formed multi-droplet configuration.
Starting from a hexagonal droplet array at $\alpha=\pi/6$ with population imbalance $N_1/N_2=6$ [\reffig{fig:n6_pi6_dynamic20ms} (a)], we apply the hysteresis protocol
\[
\alpha(t)=
\begin{cases}
	\frac{\pi}{6}\bigl(1-\tfrac{t}{20\,\mathrm{ms}}\bigr), & 0\le t\le 20\,\mathrm{ms},\\[2pt]
	0, & 20\,\mathrm{ms}< t\le 80\,\mathrm{ms},\\[2pt]
	\frac{\pi}{6}\,\tfrac{t-80\,\mathrm{ms}}{20\,\mathrm{ms}}, & 80\,\mathrm{ms}< t\le 100\,\mathrm{ms},
\end{cases}
\]
and hold $\alpha=\pi/6$ thereafter; see \reffig{fig:n6_pi6_dynamic20ms}. Throughout the cycle, the droplet number and lattice-scale arrangement remain essentially unchanged: no dislocations or vacancy/interstitial defects are nucleated within the simulated time. This indicates that the array responds \emph{elastically} to the anisotropy modulation.

As $\alpha$ is reduced to zero, individual droplets continuously evolve from flattened shapes to more rounded (elliptical) profiles; when $\alpha$ is restored, they revert to the initial flattened shapes [compare Figs.~\ref{fig:n6_pi6_dynamic20ms} (a)--(c), (c)--(e), and (e)--(g)].

The absence of plastic events, namely structural changes such as variations in droplet number, lattice constant, or defect density, implies that the energy barriers separating distinct lattice topologies exceed the work done by the slow modulation of $\alpha$ in this parameter window.
The array therefore explores a shallow valley of the energy landscape in which the primary soft degree of freedom is the intradroplet quadrupolar distortion (shape) rather than interdroplet rearrangement (positions).
This is consistent with the fact that changing $\alpha$ primarily rotates and tunes the anisotropy of the \emph{in-plane} dipolar kernel, which couples most strongly to droplet shapes (quadrupolar response), while the lattice constant is set by the balance of trap curvature and inter-droplet repulsion and is only weakly affected by moderate changes of $\alpha$.

\begin{figure*}[t]
	\begin{center}
		\includegraphics[width=\textwidth]{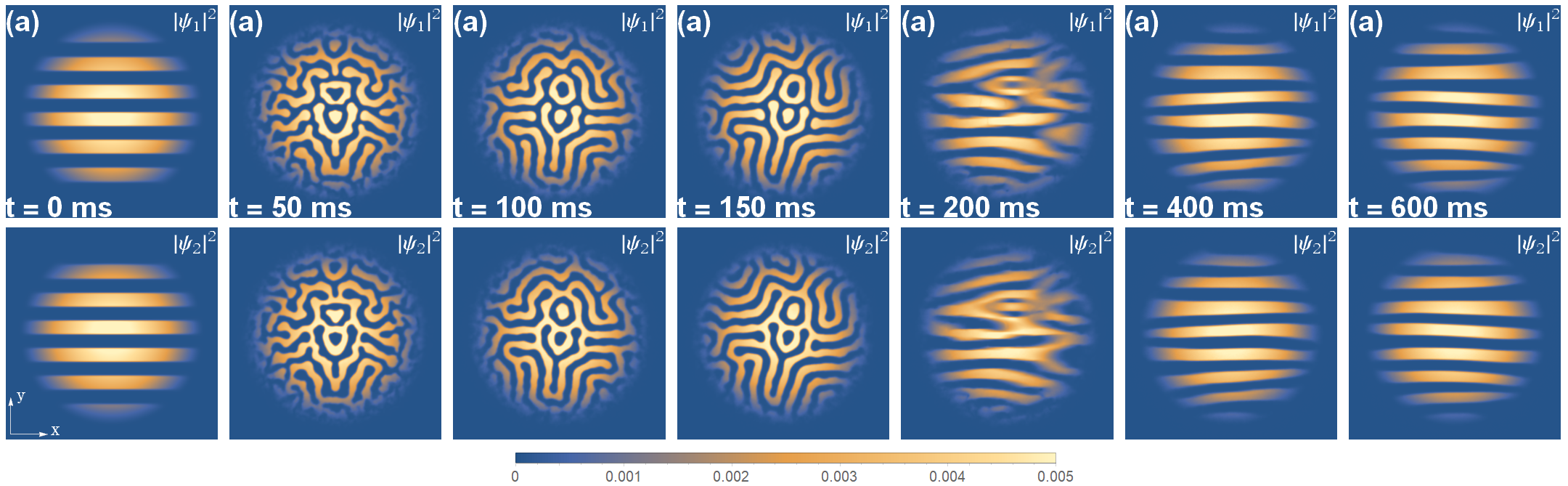}
	\end{center}\vspace{-0.5cm}
	\caption{\label{fig:stripe_hys}
		Column-density profiles $|\psi_{1}|^{2}$ and $|\psi_{2}|^{2}$ for $N_1 / N_2 = 1$ under a hysteresis protocol.
		(a) Ground state at $\alpha=\pi/3$.
		(a)–(c) The angle is ramped linearly to $0$ within $20$ ms; (c)–(e) $\alpha=0$ is held until $80$ ms; (e)–(g) the angle is ramped back to $\pi/3$ by $t=100$ ms.
		Other parameters: $a_{11}=a_{22}=a_{12}=100 a_B$, $(\omega_{\perp},\omega_{z})=2 \pi \times (100,800)$ Hz, $\mu_1 = 6 \mu_B$, and $\mu_2 = -6 \mu_B$.}
\end{figure*}

It is instructive to contrast this behavior with the stripe-to-labyrinth evolution discussed earlier.
There, reducing $\alpha$ unselects a stripe direction and activates a corrugation instability that \emph{destroys} translational order (see \reffig{fig:stripe_hys}); here, starting from a robust droplet array, the same control acts mainly as a weak anisotropic stress that \emph{deforms} but does not melt the crystal.
In other words, once droplets have formed and the lattice has annealed, the system acquires a substantial rigidity against polarization-angle cycling.

Altogether, the cycle demonstrates that the crystalline droplet array is robust under polarization hysteresis in population-imbalanced mixtures.
The response is dominated by reversible intradroplet shape distortions with negligible changes to lattice topology, providing a practical route to \emph{in situ} control of droplet ellipticity without sacrificing crystalline order.

\section{Trap-aspect-ratio effect}\label{sec:trap_ratio}

\begin{figure}[t]
	\begin{center}
		\includegraphics[width=0.48\textwidth]{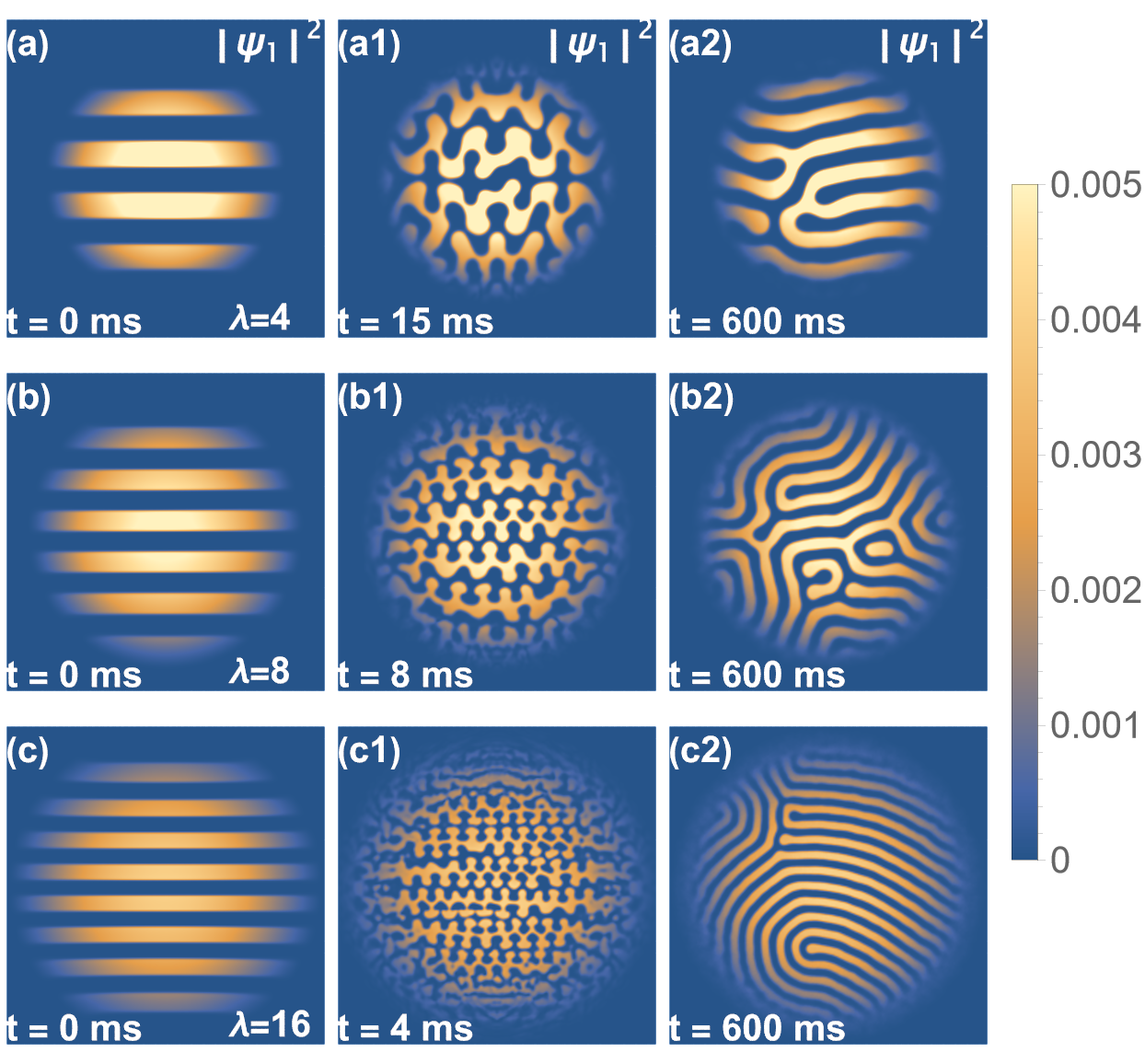}
	\end{center}\vspace{-0.5cm}
	\caption{\label{fig:n1_pi3_diffit_lambdaz}
		Column-density profiles $|\psi_{1}|^{2}$ following a quench of the polarization angle for different trap aspect ratios $\lambda=\omega_z/\omega_{\perp}$.
		(a)–(a2) $\lambda=4$, (b)–(b2) $\lambda=8$, (c)–(c2) $\lambda=16$.
		Panels (a)–(c): initial states at $\alpha=\pi/3$; at $t=0$ ms the angle is quenched to $\alpha=0$ and held during real-time evolution.
		Other parameters: $\omega_{\perp}=2 \pi \times 100$ Hz, $N_1/N_2=1$, $N_1=N_2=2 \times 10^{6}$, and $a_{11}=a_{22}=a_{12}=100 a_B$.}
\end{figure}

We now examine how the quench dynamics depend on the trap aspect ratio
$\lambda=\omega_z/\omega_{\perp}$. 
We prepare stripe ground states at $\alpha=\pi/3$ for $\lambda\in\{4,8,16\}$ with $N_1/N_2=1$ and then, at $t=0$ ms, quench the polarization angle to $\alpha=0$ while keeping $\omega_{\perp}=2 \pi \times 100$ Hz fixed; see \reffig{fig:n1_pi3_diffit_lambdaz}.
With increasing $\lambda$, the initial number of visible stripes grows [\reffig{fig:n1_pi3_diffit_lambdaz} (a)--(c)], and after the quench each stripe develops mushroom-like corrugations more rapidly [\reffig{fig:n1_pi3_diffit_lambdaz} (a1)--(c1)].
At late times, all cases evolve into labyrinthine textures, but larger $\lambda$ yields finer, more crowded labyrinths with reduced domain spacing [\reffig{fig:n1_pi3_diffit_lambdaz} (a2)--(c2)].

Tightening the axial confinement reduces the oscillator length
$l_z=\sqrt{\hbar/(m\omega_z)}=\lambda^{-1/2}\sqrt{\hbar/(m\omega_{\perp})}$, and therefore enhances the quasi-2D interaction strengths.
For contact interactions one has $g^{2\mathrm{D}}_{ij}=g_{ij}/(\sqrt{2\pi}\,l_z)$, and the in-plane dipolar kernel gains larger weight at the rotonic wave numbers through its form factor $U^{2\mathrm{D}}_d(\mathbf{k};l_z,\alpha)$.
Consequently, at larger $\lambda$, more stripe crests fit across the cloud, and a faster transverse ("mushroom") instability occurs after the quench, since the growth rate of unstable modes increases as the roton minimum softens.

Right after the quench ($\alpha: \pi/3\!\to\!0$), straight stripes first sprout transverse ripples.
The time to visible corrugation decreases monotonically with $\lambda$ [cf. \reffig{fig:n1_pi3_diffit_lambdaz} (a1) vs. (c1)].
During the intermediate stage, ripples merge into meandering filaments and short stripe segments; orientational order collapses more quickly for larger $\lambda$.
At late times, a smaller domain size shows in the labyrinth.

Because $\omega_{\perp}$ is fixed, increasing $\lambda$ at constant atom number compresses the wave function along $z$ and raises the effective 2D interaction scale, which in turn modestly increases the peak column density.
This increases the chemical potential and amplifies the contrast of the emerging patterns, aiding the quicker onset of corrugations.

The aspect ratio $\lambda$ is an experimentally clean control knob: tightening $\omega_z$ at fixed $\omega_{\perp}$ simultaneously increases the initial stripe number at $\alpha=\pi/3$, shortens the time window in which straight stripes survive after a quench to $\alpha=0$, and produces denser labyrinths at long times.
Within this regime, the observed $\lambda$-dependence provides a robust route to tune the pattern wavelength and the quench-induced instability timescale in a controlled manner.

\section{Analogy to nuclear ``pasta'' and frustrated pattern formation}\label{sec:nuclear_pasta}

Our dipolar stripes and labyrinths are closely analogous, at the level of morphology and instability routes, to the nuclear ``pasta'' phases expected in neutron-star crusts and supernova cores: rod-like and slab-like nuclear structures, cylindrical bubbles, and complex bicontinuous domains \cite{WatanabePRL2005,CaplanRMP2017}. 
In both settings, mesoscale order emerges from competing interactions that favor distinct length scales and orientations (nuclear attraction versus long-range Coulomb repulsion in pasta, and short-range contact plus anisotropic dipolar forces here), leading to frustration and self-organized textures.

Dynamically, our protocols parallel the compression-driven pathways modeled in quantum-molecular-dynamics studies of pasta formation and transitions. 
There, gentle compression converts a crystalline arrangement into rod lattices and then slabs, with intermediate defect-rich states \cite{WatanabePRL2009,WatanabePRL2005}; here, reducing the polarization tilt lowers the effective transverse stiffness and drives a corrugation that severs stripes into meanders and labyrinths. 
The shared phenomenology (instability of a nearly one-dimensional order---rods/stripes---transient defect proliferation, and eventual domain labyrinths) suggests a common organizing principle: a soft mode near a characteristic wave number that becomes unstable as a control parameter (density or tilt) crosses a threshold.

There are, however, important differences. 
Our textures are trap-bound condensate patterns in a quasi-two-dimensional, zero-temperature mean-field regime where beyond-mean-field corrections are negligible and droplets are not self-bound, whereas nuclear pasta forms in charge-neutral matter embedded in an electron gas and is stabilized by nuclear saturation and Coulomb frustration \cite{WilsonPRL1983,YamadaPTP1984,CaplanRMP2017}. 
Accordingly, the quantitative scales and conservation laws differ, but the qualitative taxonomy of patterns and the role of frustrated interactions are strikingly similar, providing a useful cross-disciplinary lens for interpreting our stripe–labyrinth transitions and droplet arrays.

\section{Conclusions}\label{sec:conclusion}

We have mapped non-equilibrium pattern formation in a two-component dipolar BEC with opposite polarizations by tuning the polarization angle $\alpha$ and the trap aspect ratio $\lambda=\omega_z/\omega_\perp$. At finite $\alpha$ the system supports stripe order; a sudden quench to $\alpha=0$ excites a characteristic mushroom-like corrugation that destroys translational order and yields labyrinthine textures, whereas a slow linear ramp produces transient curved stripes before converging to labyrinths, consistent with Ref. \cite{ArazoGallem2023}. Population imbalance biases the dynamics: the minority component preferentially fragments into a stable array of droplets after quenches or ramps while the majority remains comparatively diffuse on accessible timescales. Once crystalline arrays are formed, they exhibit robust hysteresis---cycling $\alpha$ primarily induces reversible shape changes with essentially unchanged lattice topology. The trap aspect ratio further tunes both the initial stripe count and the instability timescale; increasing $\lambda$ enhances the quasi-2D character, accelerates the corrugation, and produces denser labyrinths at long times; see also Ref. \cite{PhysRevLett.107.150403}.

A unified picture emerges in terms of roton-assisted instabilities: reducing $\alpha$ lowers the transverse stiffness, broadens the unstable band, and seeds the observed corrugations and domain proliferation. These trends are directly testable via image-based diagnostics within the few-$100$ ms windows used here. Our analysis operates in a zero-temperature, quasi-2D mean-field regime where beyond-mean-field corrections are negligible for our densities; accordingly, the droplets reported here are not self-bound in free space. The required controls, namely magnetic-field tilt (setting $\alpha$), trap geometry (setting $\lambda$), and atom-number imbalance, are standard and offer complementary knobs for selecting and steering patterns.

Beyond these dipolar-gas specifics, the morphologies and their routes of formation closely echo the ``nuclear pasta'' patterns (rods, slabs, and defect-rich intermediates) predicted and simulated for neutron-star crusts and supernova cores \cite{WatanabePRL2005,WatanabePRL2009}. The microscopic interactions and constraints differ (contact and dipolar forces in a trapped quantum gas versus nuclear attraction and Coulomb frustration in nearly charge-neutral matter), yet both systems display pattern selection driven by competing interactions and a soft mode near a characteristic length scale. Our results thus place dipolar BECs alongside nuclear pasta as a clean, controllable platform for studying frustrated mesoscale order. Within our quasi-two-dimensional mean-field regime, beyond-mean-field effects are negligible and droplets are not self-bound. The demonstrated controls---polarization angle, trap geometry, and population imbalance---provide a practical route toward a pattern-phase diagram for dipolar mixtures and motivate quantitative follow-ups, including finite-temperature extensions and spectroscopy of the roton-driven instability.

\begin{acknowledgments}
K.-T.X. was supported by the MOST of China (Grant No. G2022181023L) and NUAA (Grant No. YAT22005, No. 2023YJXGG-C32 and No. XCA2405004). W.B. was supported by NUAA (Grant No. 202510287091Z). G.W. was supported by National Natural Science Foundation of China (Grant No. 12375039). 
\end{acknowledgments}

\bibliography{dBECbinary}

\end{document}